\documentclass[twocolumn]{aastex63}
\usepackage{graphicx}
\usepackage{bm}
\usepackage{amssymb}
\usepackage{amsmath}
\usepackage{units}
\usepackage{array}
\usepackage{lineno}
\usepackage{ulem}
\received{November 10, 2021}
\revised{April 19, 2022}
\accepted{May 2, 2022}
\published{June 13, 2022}
\submitjournal{ApJ}

\shorttitle{How binaries accrete}
\shortauthors{Tiede et al.}

\graphicspath{{./}{figures/}}

\begin{document}
\title{How binaries accrete: hydrodynamics simulations with passive tracer particles}

\author[0000-0002-3820-2404]{Christopher Tiede}
\affiliation{Center for Cosmology and Particle Physics, Physics Department, New York University, New York, NY 10003, USA}

\author[0000-0002-1895-6516]{Jonathan Zrake}
\affiliation{Department of Physics and Astronomy, Clemson University, Clemson, SC 29634, USA}

\author[0000-0002-0106-9013]{Andrew MacFadyen}
\affiliation{Center for Cosmology and Particle Physics, Physics Department, New York University, New York, NY 10003, USA}

\author[0000-0003-3633-5403]{Zolt\'an Haiman}
\affiliation{Department of Astronomy, Columbia University, New York, NY 10027, USA}

%
% =============================================================================
\begin{abstract}
Linear analysis of gas flows around orbiting binaries suggests that a centrifugal barrier ought to clear a low-density cavity around the binary and inhibit mass transfer onto it. 
Modern hydrodynamics simulations have confirmed the low-density cavity, but show that any mass flowing from large scales into the circumbinary disk is eventually transferred onto the binary components. 
Even though many numerical studies confirm this picture, it is still not understood precisely how gas parcels overcome the centrifugal barrier and ultimately accrete. We present a detailed analysis of the binary accretion process, using an accurate prescription for evolving grid-based hydrodynamics with Lagrangian tracer particles that track the trajectories of individual gas parcels.
We find that binary accretion can be described in four phases: (1) gas is viscously transported through the circumbinary disk up to the centrifugal barrier at the cavity wall, (2) the cavity wall is tidally distorted into accretion streams consisting of near-ballistic gas parcels on eccentric orbits, (3) the portion of each stream moving inwards of an ``accretion horizon'' radius $\bar r \simeq a$---the radius beyond which no material is returned to the cavity wall---becomes bound to a minidisk orbiting an individual binary component, and (4) the minidisk gas accretes onto the binary component through the combined effect of viscous and tidal stresses.
\end{abstract}
%

%
% =============================================================================
%

%
% =============================================================================
\section{Introduction} \label{sec:intro}
%

% Uncomment to discover the column and text width in inches:
% \makeatletter
% \def\convertto#1#2{\strip@pt\dimexpr #2*65536/\number\dimexpr 1#1}
% \makeatother
% Text width: \convertto{in}{\the\textwidth}, Column width: \convertto{in}{\the\columnwidth}.

Circumbinary disks are frequent evolutionary accessories to binaries spanning celestial scales.  
They are observed or expected from small scales around planet-moon systems \citep{PDS70c}, star-planet binaries in protoplanetary nebulae \citep{Ward1997, Kley2012}, and young stellar binaries \citep{DQTau, McCabe2002, Krist2005, Orosz2012} up to much larger scales around massive black hole pairs \citep{al96, Armitage2002, Milosavljevic2005}. 
The problem of the dynamics and observable signatures of such circumbinary disks has been a central issue in astrophysics for decades because the interaction between the disk and central binary is vital for both the evolution of embedded moons, planets, binary stars, or black holes as well as the identification of such binaries in astronomical surveys.  
Specifically, it is important to understand the complex flow of material around compact binaries near the central regions of the disk--well within the self-gravitating radius--in order to develop detailed understandings of the binary mass accretion rate, the observational signatures of the disk itself, and the effects of accretion and gravitational forces on the orbital evolution of the central components.

The problem of mass flow in the central regions of a circumbinary disk was at first primarily studied in the context of low-mass-ratio planets in protoplanetary disks.  
In such a situation the tidal torques launched from the low-mass satellite 
perturb a standard Keplerian accretion disk, expelling material from the corotation region and carving out an annular gap in the orbital path of the satellite \citep{LP86, GT80, al94}.  
As the mass-ratio of the satellite is increased, however, the annular gap widens. Beyond a mass-ratio $q\sim 0.04$ \citep{Dorazio2016}, the tidal torques drive a morphological transition in the disk whereby the binary depletes material from the entire central region manufacturing an evacuated cavity of radius approximately double the semi-major axis of the binary \citep[e.g.][]{al96, Escala2005, MacFadyen2008, Dorazio2013, Farris15, MML17}.  

In this case of large mass-ratio ($q \gtrsim 0.04$), cavity carving binaries, this understanding is mostly empirical. Original analytic study of the dynamics of gas disks around binaries of similar mass (e.g. $q \gtrsim 0.1$) predicted that the tidal torques from these large-$q$ binaries would act as a dam against the accretion flow, suppressing and possibly even shutting off gas accretion onto the binary components \citep{Pringle1991, Milosavljevic2005, LiuShapiro2010,Kocsis+2012a,Kocsis+2012b}.  
For observational purposes, such a result could markedly diminish the feasibility of observing compact dual-AGN and late-stage, pre-gravitational-inspiral massive black hole binaries. 
However, these studies considered the one dimensional case
assuming axisymmetry. 

Numerical simulations of this problem revealed that while the tidal distortions from the binary do in fact carve out a depleted central cavity, the cavity and its associated features are far from axisymmetric: The cavity is found to be lopsided and to precess at approximately the analytic quadrupolar frequency, there exists an $m=1$ density feature, or ``lump'', that orbits that binary at $\sim 5$ times the binary orbital period, each binary component forms its own accretion disk, termed a ``minidisk," and unstable stream-like structures form on dynamical times, delivering material from the wall of the outer-cavity onto the binary minidisks \citep{MacFadyen2008, Cuadra2009, Shi+2012, Farris15, Shi2015, Yike17, MML19, Ragusa2020, Munoz2020}. 
Of primary importance, a series of these numerical studies measured the rate of gas accretion onto the central binary and found that the binary accretion rate is nearly identical to that expected for a single object embedded in a Keplerian accretion disk \citep{Roedig2012, Shi+2012, Dorazio2013, Farris15}.  
However, there is work showing that the accretion rate can be sensitive to disk parameters such as the thickness of the disk \citep{Ragusa+2016, Tiede2020}.

Despite this growing empirical consensus regarding the dynamics of accretion flows in the vicinity of astrophysical binaries, the literature is lacking a physical explanation describing how gas is able to penetrate the barrier of the rotating binary's Roche potential, cross the evacuated cavity, and accrete onto a binary component. 
\cite{Shi2015} showed that there are specific orbital parameters that result in the dynamical accretion of a parcel of gas and posit that such phase-space coordinates are achieved by the shock deflection of rejected stream material as it impacts the cavity wall (although this remains to be demonstrated).
The goal of this paper is to develop a physical description for how fluid elements
travel from the outer disk, into accretion streams, and onto minidisks through which they are ultimately accreted; namely, to describe how binaries accrete.  
To do so, we simulated the simplest case of a circular, equal-mass binary accreting from a thin disk at high resolution in two dimensions with $\mathcal{O}(10^6)$ passive tracer particles so as to follow and analyze the accretion histories of fluid elements in the disk.
We focus first on understanding this problem for 2D, isothermal disks and leave the effects of General Relativity, magnetic fields, and radiation to future investigation.

This paper is organized as follows. In \S~\ref{sec:simulations} we describe the details of our computational methodology and setup.  In \S~\ref{sec:results} we present the results of our simulations and the analysis. Finally in \S~\ref{sec:summary} we summarize our main results and discuss some of their implications.

%
% =============================================================================
\section{Numerical Methods} \label{sec:simulations}

In this section, we detail the numerical tools and experiments used to explore the question of how binaries accrete.  
All simulations were performed using an upgraded version of the code \texttt{Mara3} \citep{Mara, Tiede2020, Zrake2020} written in Rust (\texttt{Mara-F3O}).  

\subsection{Simulation setup}
\label{sec:setup}

\texttt{Mara-F30} solves the vertically-averaged Navier-Stokes equations
\begin{align}
    \partial_t \Sigma &+ \nabla \cdot (\Sigma \boldsymbol v) = \dot \Sigma_{\rm sink}
  \label{eqn:continuity} \\
    \partial_t (\Sigma \boldsymbol v) &+ \nabla \cdot \left( \Sigma \boldsymbol v \boldsymbol v + P \mathbf I - \boldsymbol \tau \right) = \dot \Sigma_{\rm sink} \boldsymbol v + \boldsymbol F_{\rm g}
  \label{eqn:momentum}
\end{align}
via a finite volume Godunov scheme in Cartesian coordinates.  
In equations \ref{eqn:continuity} and \ref{eqn:momentum}, $\Sigma$ denotes the vertically integrated surface density of the disk, $\boldsymbol v$ is the gas velocity vector, and $P = \Sigma\, c_s^2$ is the vertically integrated pressure in an isothermal disk. 
In equation \ref{eqn:momentum}, $\boldsymbol \tau$ is the viscous stress tensor, $\dot \Sigma_{\rm sink}$ is a mass removal term meant to model the accretion of material onto each component, and $\boldsymbol F_g = -\Sigma \nabla \phi_g$ is the vertically integrated gravitational force density.  
The gravitational potential is 
\begin{align}
    \phi_g = -\frac{GM_1}{(r_1^2 + r_s^2)^{1/2}} -\frac{GM_2}{(r_2^2 + r_s^2)^{1/2}}
\end{align}
with $r_1$ and $r_2$ the distances from the respective binary component, and $r_s$ the gravitational softening length to account for the vertical averaging of the gravitational force and to prevent its divergence.  
$r_s$ is nominally chosen to be 5\% of the binary semi-major axis.  
The disk is chosen to have scale-height $(h / r) = 0.1$ with isothermal equation of state
\begin{align}
    c_s = \left( \frac{h}{r}\right) \, \sqrt{-\phi_g} \ ,
\label{eq:eos}
\end{align}
and the viscous stress tensor is given as
\begin{align}
    \tau_{ij} = \nu \Sigma \left( \partial_i v_j + \partial_j v_i - \delta_{ij} \partial_k v_k \right) \ .
\end{align}
The viscosity is set to be constant as $\nu = 10^{-3}$. 

In order to model the sub-grid accretion of material onto the central objects, we employ a ``standard'' mass sink \citep{Farris15, Yike17, MML19, Moody19, Tiede2020, Duffell2020} of radius $r_{\rm sink}$ and removal timescale $t_{\rm sink}$
\begin{align}
    \frac{\dot \Sigma_{\rm sink}}{\Sigma} = -\left( e^{-(r_1 / r_{\rm sink})^6} + e^{-(r_2 / r_{\rm sink})^6} \right)\,t_{\rm sink}^{-1} \ .
\end{align}
$r_{\rm sink}$ is taken as the gravitational softening length $r_s$ and the removal timescale is chosen in the marginally fast limit, 
$t_{\rm sink}^{-1} = 10 \,\Omega_b$ (where $\Omega_b$ is the binary's orbital frequency ).
The choice of $t_{\rm sink}$ does not significantly alter the results for circular orbits \citep{Moody19, RyanWS2021, Zrake2022}, but for robustness we include a brief exploration in the slow sink limit with $t^{-1}_{\rm sink} = \Omega_b$ in Section \ref{sec:accretion-horizon}.
Recently, \cite{Dempsey2020} and \cite{Dittmann+Ryan2021} have suggested that using so-called ``torque-free'' sinks can reduce sensitivity to sink parameters, but the latter similarly found negligible variations for equal mass binaries. 
In future work we intend to explore the effect of such torque-free sinks on this paper's results.

The disk is initialized with peak surface density at $r_d = 4a$ and a mildly depleted cavity region 
\begin{align}
    \Sigma = \Sigma_0\,e^{-(r / r_d - 1)^2}
    \label{eq:sigma0}
\end{align}
with pressure-corrected Keplerian velocity
\begin{align}
    \boldsymbol v = \left( \frac{GM}{r} + \frac{r}{\Sigma}\frac{\partial P}{\partial r} \right)^{1/2}\,\hat \phi \ .
    \label{eq:v0}
\end{align}
This represents a steady-state solution to equations \ref{eqn:continuity} and \ref{eqn:momentum} with zero viscosity and a single, central object of mass $M = M_1 + M_2$. 

For simplicity, in this paper, we only consider equal-mass binaries fixed on a circular orbit and assume that the disk mass $M_d$ is much less than that of the binary $M_d \ll M$ such that the Toomre parameter $Q \sim (h / r) (M / M_d) \gg 1$. In this way, $\Sigma_0$ is arbitrary, and we can ignore the disk's self-gravity. 
Further, the assumption of fixed circular orbits has been supported by evidence that near-circular orbits (of eccentricity, $\epsilon \lesssim 0.1$) are driven towards the circular limit, retaining their negligible eccentricities \citep{MML19, Zrake2020}. 
The simulation domain extends out to $15\,a$ in each Cartesian direction, and the number of zones is selected to give a grid resolution of $\Delta x_{\rm cell} \approx 0.015\,a$.

\subsection{Tracer implementation}
\begin{figure*}[t!]
\includegraphics{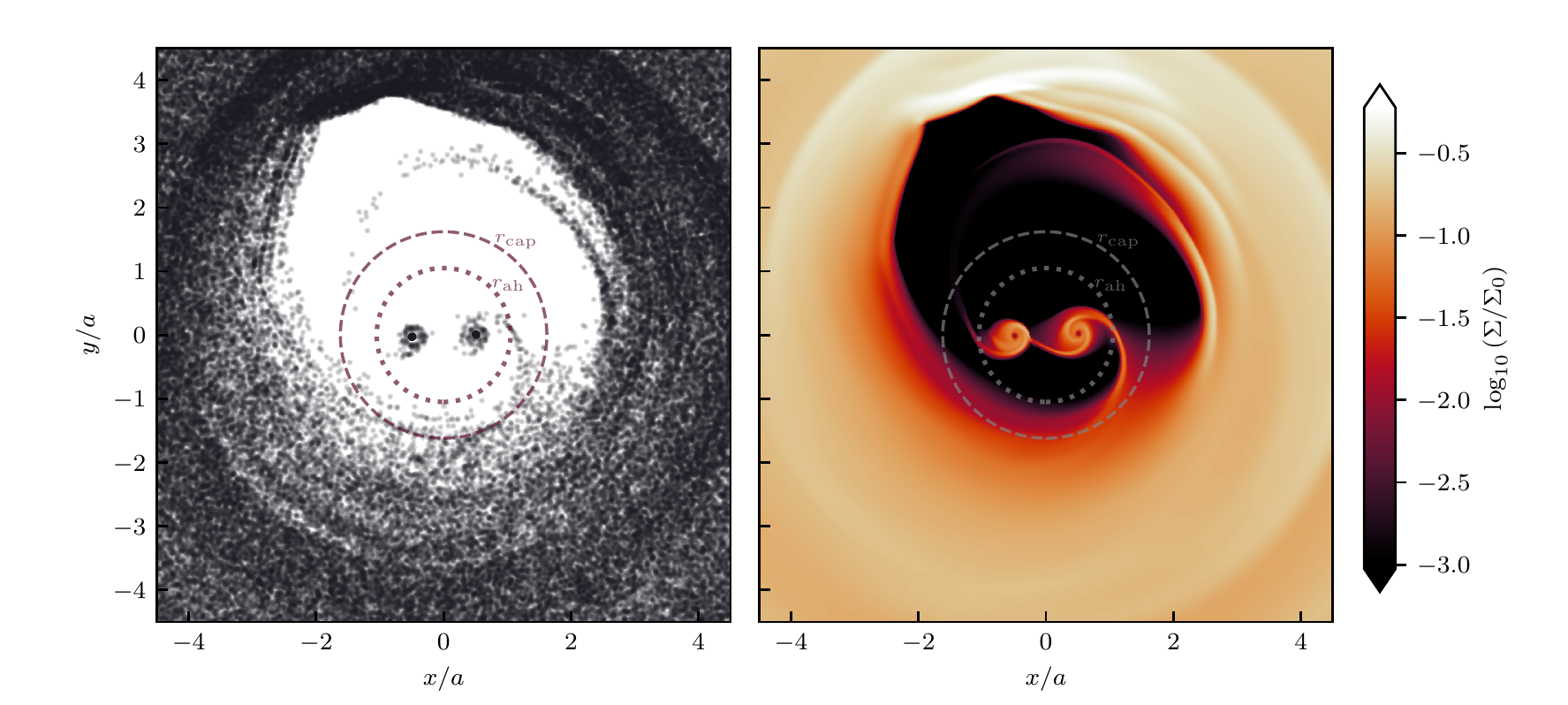}
\caption{Tracer particle positions (left) and disk surface density profile (right) in the fluid after 650 orbits. Grey circles in the right panel are included for future reference: the inner dotted circle for $r_{\rm ah} \sim 1.05\,a$ (\S \ref{sec:accretion-horizon}), and the outer dashed one for $r_{\rm cap} \sim 1.6\,a$ (\S \ref{sec:stream-formation}).}
\label{fig:reliefs}
\end{figure*}

A major drawback of Eulerian schemes is that the fluid is described by the evolution of fields at fixed spatial locations and the past history of individual fluid elements is not tracked.
One solution to this is Lagrangian smoothed particle hydrodynamics (SPH) schemes that discretize the fluid into particles that are integrated forward via derived physical fields \citep[e.g.][]{GM77, Lucy77, Monaghan1992, Price2012}, but these struggle to resolve the inner most regions of the binary accretion flow \citep{Ragusa+2016, Ragusa2020, Heath+Nixon2020}.
A middle ground between these two is to introduce passive \emph{tracer particles} into conservative Eulerian schemes \citep[e.g.][]{Ensslin2002, Dubey2012, Dubois2012}.  
These tracer particles have no mass and simply advect along with the fluid flow, but they allow one to follow hydrodynamic histories of fluid elements in a Lagrangian description of the flow.

The most common type of tracer particle, as described, is a velocity field tracer.  
These tracers compute an estimate of the local fluid velocity and integrate forward in time according to the timestepping procedure of the Eulerian solution.  
Typically this velocity is calculated either by sampling the nearest cell velocity or via higher-order interpolation schemes (although it has been demonstrated that this does not significantly affect results;
\citealt{Federrath2008, Vazza2010, Konstandin2012}).
One demonstrated drawback of choosing the tracer velocity from the reconstructed velocity field is that it can lead to over/under-densities at convergence/divergence points in the fluid. 
This is because two velocity field tracers can be arbitrarily close on two separate sides of a cell interface, and even though they are at nearly identical locations in the fluid, drawing their velocities from the reconstructed fields can imbue them with meaningfully different velocities \citep{AREPO-tracers}.

In order to overcome the mismatch between the mass flow implied by the reconstructed velocity field and the actual flow as determined by the Riemann solutions across each interface, we choose the velocity for tracer $j$ as a linear combination of the velocities associated with the flux returned by the Riemann solver at each cell interface $\boldsymbol v$($\boldsymbol F^{\rm HLL}_{i \pm 1/2}$),
\begin{align}
    \boldsymbol v_j = \left( 1 - \frac{\delta x_{ij}}{\Delta x_i} \right) \, \boldsymbol v ( \boldsymbol F^{\rm HLL}_{i - 1/2} ) + \frac{\delta x_{ij}}{\Delta x_i} \, \boldsymbol v ( \boldsymbol F^{\rm HLL}_{i + 1/2} ) 
    \label{eq:tracers}
\end{align}
with $\Delta x_i$ the cell width and $\delta x_{ij}$ the distance of tracer $j$ from the cell face at $i - 1/2$.
In this way, the tracer velocity is chosen to directly reflect the mass flow at each simulation timestep as determined by the Riemann solver.
Given that the tracers can accurately follow the mass flow in the disk, all other instantaneous hydrodynamic quantities (e.g. density or angular momentum) can be queried at any point in time in order to recreate the hydrodynamic history of a given mass parcel in a Lagrangian description of the flow.  Tests on the reliability of the tracers in this regard are presented in the Appendix.

Computationally, each tracer is defined solely by its unique ID and its current coordinates to keep it as light-weight as possible.  
The tracers are updated every time step with the same second-order Runge-Kutta scheme as the fluid, and any information about the local fluid state is recorded at each tracer data output.  
We chose the number of tracers to be comparable to the number of cells $N_{\rm tracers} \approx 2.3 \times 10^6 \sim N_{\rm cells}$.
For all tracer results presented in this paper, tracer data was output every
$0.02\,P_b$.  
In this way, we are able to construct time series 
for given fluid elements, and a Lagrangian description of their flows in post-processing.
\subsection{The purely gravitational problem}
In addition to the ability to track the Lagrangian histories of fluid parcels, the use of tracer particles also provides us with the 4-dimensional phase-space coordinates of said fluid elements at any tracer output time in the simulation.  This enables the ability to compare the full hydrodynamic evolution of a fluid element with purely gravitational evolution.

In the purely gravitational problem -- the circular restricted 3-body problem (cr3bp) -- the only constant of motion is a particle's Jacobi constant
\begin{align}
    c_j = -2\,U - {\tilde v}^2
 \label{eq:cj}
\end{align}
where $U = \phi_g - \left( \boldsymbol \Omega_b \times \boldsymbol r \right)^2$ is the Roche potential\footnote{Note that sometimes $U$ is defined as the negative of the Roche potential to eliminate the leading minus sign in Equation \ref{eq:cj}} and $\tilde{v}$ is the particle velocity in the binary orbital frame. 
In addition to being a constant of motion, $c_j$ defines restricted regions in the binary orbital plane separated by so-called zero-velocity-curves (ZVCs) set by the condition $c_j < -2\,U$.  
There exists a family of such ZVCs that connect and divide the binary orbital plane into $\geq 2$ distinct topological regions. 
For equal mass binaries, the curve in this set with the smallest value of $c_j$ is that which goes through the $L2$ and $L3$ Lagrange points.  
This is given as $c_j^{\rm crit} = 3.64 \,\Omega_b^2\,a^2$, and gravitational orbits with $c_j > c_j^{\rm crit}$ are topologically confined to either the outer disk or to one of the minidisks (see, e.g., Fig.~2 in \citealt{Dorazio2016}) -- i.e. they are gravitationally incapable of moving from the outer disk, across the cavity, and onto a minidisk without the assistance of other sources like pressure or viscosity.  On the other hand, particles with $c_j < c_j^{\rm crit}$ are dynamically allowed to cross the cavity ballistically.

When integrating orbits, we used an adaptive Dormand-Prince, fifth-order Runge-Kutta method that explicitly conserved $c_j$ to fractional order $10^{-6}$ or better. 

%
% =============================================================================
\section{Results and Discussion} \label{sec:results}
\begin{figure}[t!]
\includegraphics{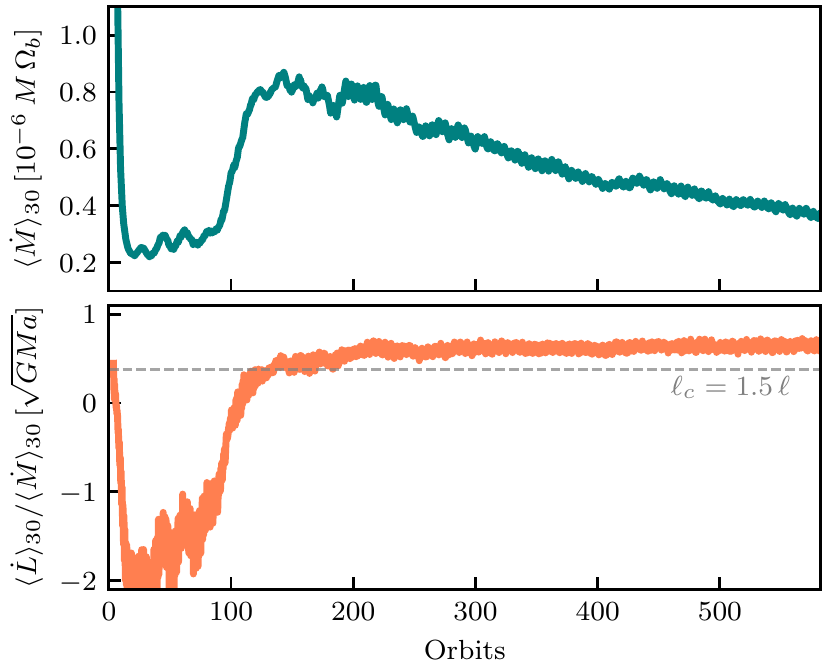}
\caption{Accretion rate and change in binary angular momentum per unit accreted mass. The disk achieves a quasi-steady state (e.g. approximately constant angular momentum delivery to binary) after $\sim 200$ orbits.}
\label{fig:qsteady}
\end{figure}

We ran our simulations for more than a viscous time at the cavity wall ($r = 2.5\,a$), $\sim 700$ orbits, before recording tracer trajectories, so that the disk has relaxed into a quasi-steady state. The right panel of Fig. \ref{fig:reliefs} shows the surface density of the disk after 650 orbits. The left panel shows the distribution of tracers embedded in the disk.
We see that the tracers follow the flow morphology, as reflected in the minidisks, the cavity shape and structure\footnote{We note that the kink-like features at the cavity edge occur as the result of shock interactions in the cavity wall and can similarly be observed in other high-resolution binary simulations \citep[e.g.][]{Munoz19, Duffell2020, Dorazio2021, Dittmann+Ryan2021}.}, the streams, and the density waves in the outer disk.

We define the quasi-steady configuration as one in which the specific angular momentum imparted to the binary per unit accreted mass $\ell_0 \equiv \dot L / \dot M$ is approximately constant. 
A 30 orbit moving average of this quantity is shown in the bottom panel of Figure \ref{fig:qsteady} alongside the time-averaged binary accretion rate in the top panel\footnote{See \citealt{Tiede2020} for details on the quasi-steady condition and specifics on how these quantities are calculated.}, $\dot M$.
In contrast with $\ell_0$, $\dot M$ possesses some secular evolution as the finite CBD is slowly depleted, but it has been demonstrated that this does not significantly affect the angular momentum transfer \citep{Munoz19}. 
We observe the startup phase for the first $\sim 200$ orbits, after which the disk settles into said quasi-steady configuration.
We also find, consistent with other recent studies \citep[e.g.][]{MML19, Moody19, Tiede2020}, that the equilibrium value of the accretion eigenvalue $\ell_0$ for $(h/r) = 0.1$ disks is greater than the critical value $\ell_c = 1.5 \,\ell$, meaning that the binary experiences orbital softening, $\dot a > 0$. $\ell_0 = \ell_c$ would mean that the binary receives just enough angular momentum to balance orbital hardening caused by the addition of mass to the components.

\subsection{Qualitative picture}
\label{sec:qualpic}

Mass is transported inwards through the CBD by the (effective) viscous stress. Gas parcels begin to feel the influence of the binary in the range $3.5 a - 8 a$, in the form of random deflections from outward-propagating pressure waves, launched from the CBD inner edge at $r \simeq 2.5a$. Indeed, gas motions around the CBD wall $2.0a - 3.5a$ are highly unsteady due to the strong tidal influence of the binary (\S \ref{sec:accretion-histories}). Fluid elements experience strong orbital radialization in this range, developing increasingly eccentric orbits as they move inwards. The CBD wall itself is highly eccentric with $e \simeq 0.3$ (\S \ref{sec:stream-formation}). These observations are consistent with established picture of circumbinary accretion.

What has remained unclear until now is how material from the cavity wall loses enough angular momentum to enter the low-density cavity around the binary, and ultimately join one of the minidisks. The tracer particles enable us to measure precise trajectories of the gas parcels, and answer this question directly. We observe that the gas flow into the cavity is generally in the form of a narrow, fast-moving stream connecting the cavity wall to a minidisk (see Figure \ref{fig:reliefs}). An important clue as to the dynamics of the accretion streams is that they form and dissolve twice during each binary orbit. In \S \ref{sec:stream-formation} we will show that the stream formation is the result of tidal stressing from the binary; it is a uniquely gravitational (as opposed to hydrodynamical) process.

\begin{figure}
\includegraphics{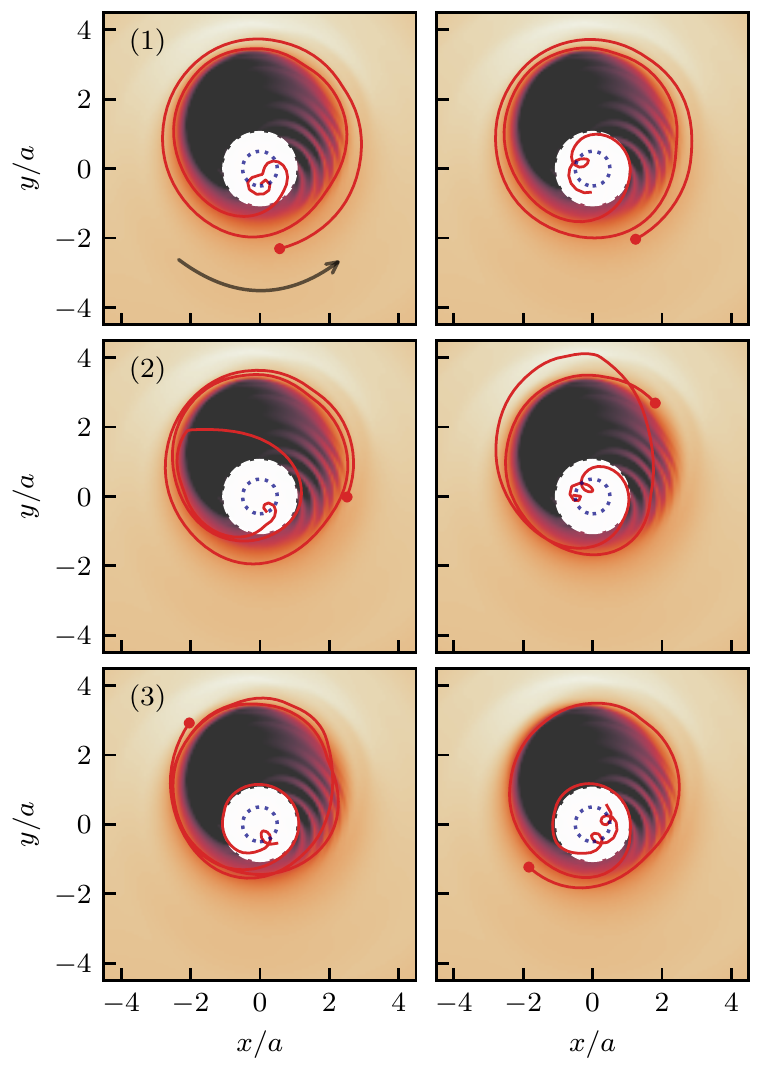}
\caption{Accretion trajectories in the observer frame representative of the scenarios described in Section \ref{sec:qualpic}. The top row (1) are examples of fluid elements that are immediately deposited from their stream onto a minidisk; the middle row (2) are examples of those that are flung back to the cavity wall (at least once) and accrete later; and the bottom row (3) are examples of those that spend $\sim 1$ orbit hovering around a Lagrange point before ultimately joining a minidisk. The underlying map is the time averaged density field in the observer frame with the same color-scale as Figure \ref{fig:reliefs}. The central region is removed for clarity, but the dotted circle shows the binary orbit location.  Each trajectory begins at the red dot and travels counterclockwise (prograde around the binary) as indicated by the arrow in the top left panel.}
\label{fig:trajectories}
\end{figure}

We find (\S\S \ref{sec:accretion-horizon} and \ref{sec:minidisk-capture}) that after being swept into a stream, a fluid element has one of two fates: (1) it directly enters one of the binary minidisks, in less than a binary orbit following the stream formation, or (2) it is flung back to the cavity wall. Two-dimensional trajectories illustrating each scenario are shown in Fig. \ref{fig:trajectories}, where the top and middle rows contain examples of Scenarios 1 and 2 respectively\footnote{The appearance of spiral structures or a wave like feature extending from this excised region is the result of the non-physical aliasing of the time-variable stream structures since our snapshots are taken 10 times per orbit.}. 
Fluid elements returning to the cavity wall (Scenario 2) are often swept into another stream within the next few binary orbits (as shown in the middle row of Figure \ref{fig:trajectories}), but others can return to the CBD wall for tens or even hundreds of orbits. A small subset of tracer particles from Scenario 1 hover near a Lagrange point for $\sim 1$ orbit before falling onto a minidisk; examples are shown in the bottom panel of Fig. \ref{fig:trajectories}.

The accretion streams penetrate inwards and temporarily connect to one of the minidisks. The gas parcels near the leading edge of the stream are decelerated either by an accretion shock or pressure wave, and join the outer portion of that minidisk, while the gas parcels toward the trailing edge of the stream are rejected and rejoin the CBD. We show in \S \ref{sec:minidisk-capture} that some fluid elements are absorbed into the minidisk on which they are initially decelerated, while others are transferred to the other minidisk.

We can thus summarize the process of circumbinary accretion in three stages: (1) inward viscous transport, (2) tidal deformation, and (3) collisions with minidisks. The following sections quantify these processes in depth, based on data we have gathered from our hydrodynamics simulations with tracer particles.

\subsection{The accretion horizon} \label{sec:accretion-horizon}
\begin{figure}[t!]
\includegraphics{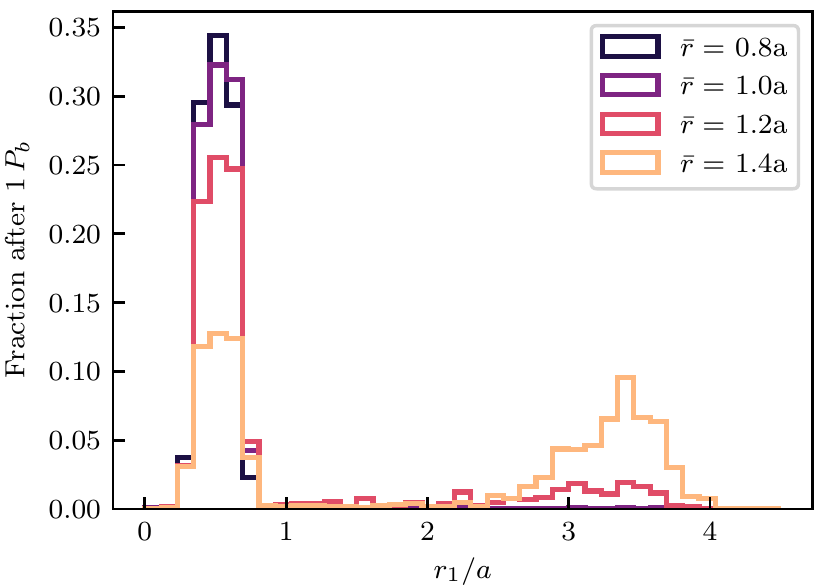}
\caption{
Distributions of tracer radii 1 orbit after a close approach to the binary.  The close approaches are defined as the time at which a given tracer eneters inside the radius $r < \bar r$.  As $\bar r$ is lowered, we see that the number of tracers returned to the cavity wall ($r \gtrsim 2 \, a$) steadily declines and eventually shuts off for $\bar r \lesssim 1.2\,a $.
}
\label{fig:rbar}
\end{figure}
\begin{figure}[h!]
\includegraphics{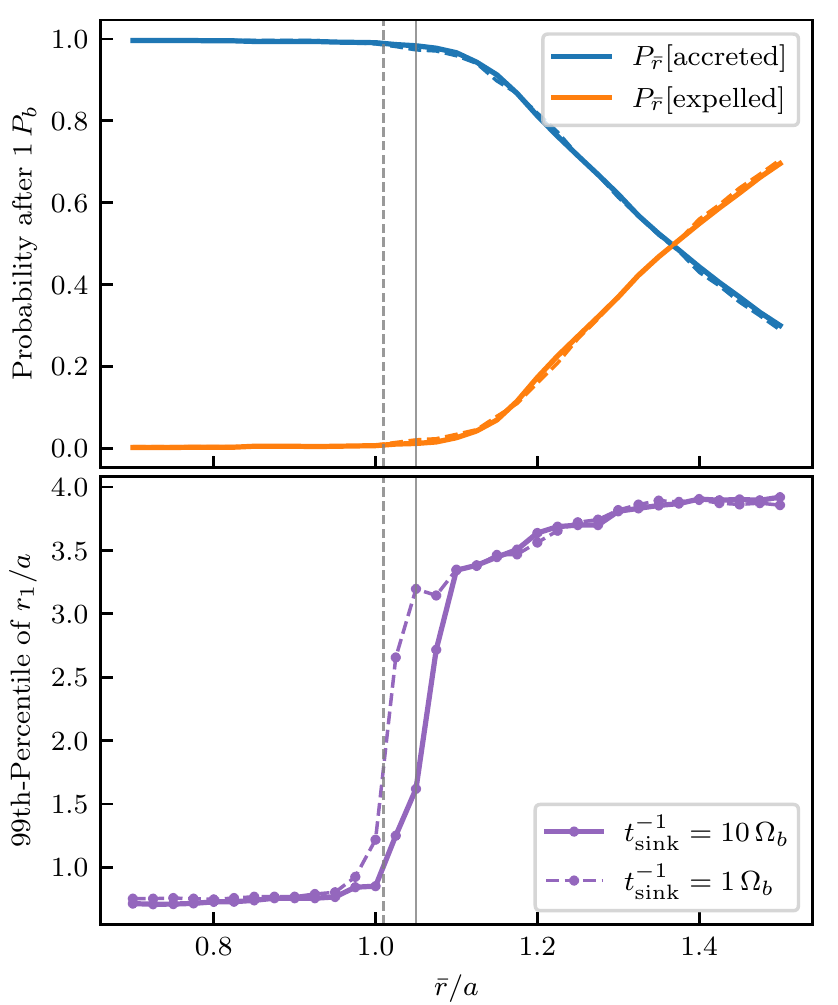}
\caption{
(Top) Probability of accretion given that a tracer has crossed inside of $\bar r$ integrated from the distributions in Figure \ref{fig:rbar}, and (Bottom) the 99th-percntile radii from said distributions as functions of the approach-radius $\bar r$ (solid lines). 
The solid vertical line indicates the accretion horizon, $r_{\rm ah} \approx 1.05\,a$.
The dotted lines show the same result for $t_{\rm sink}$ in the slow-sink limit. In this limit the horizon is shifted by a few percent, $\approx 1.01\,a$.
}
\label{fig:noreturn}
\end{figure}
\begin{figure*}[t!]
\includegraphics{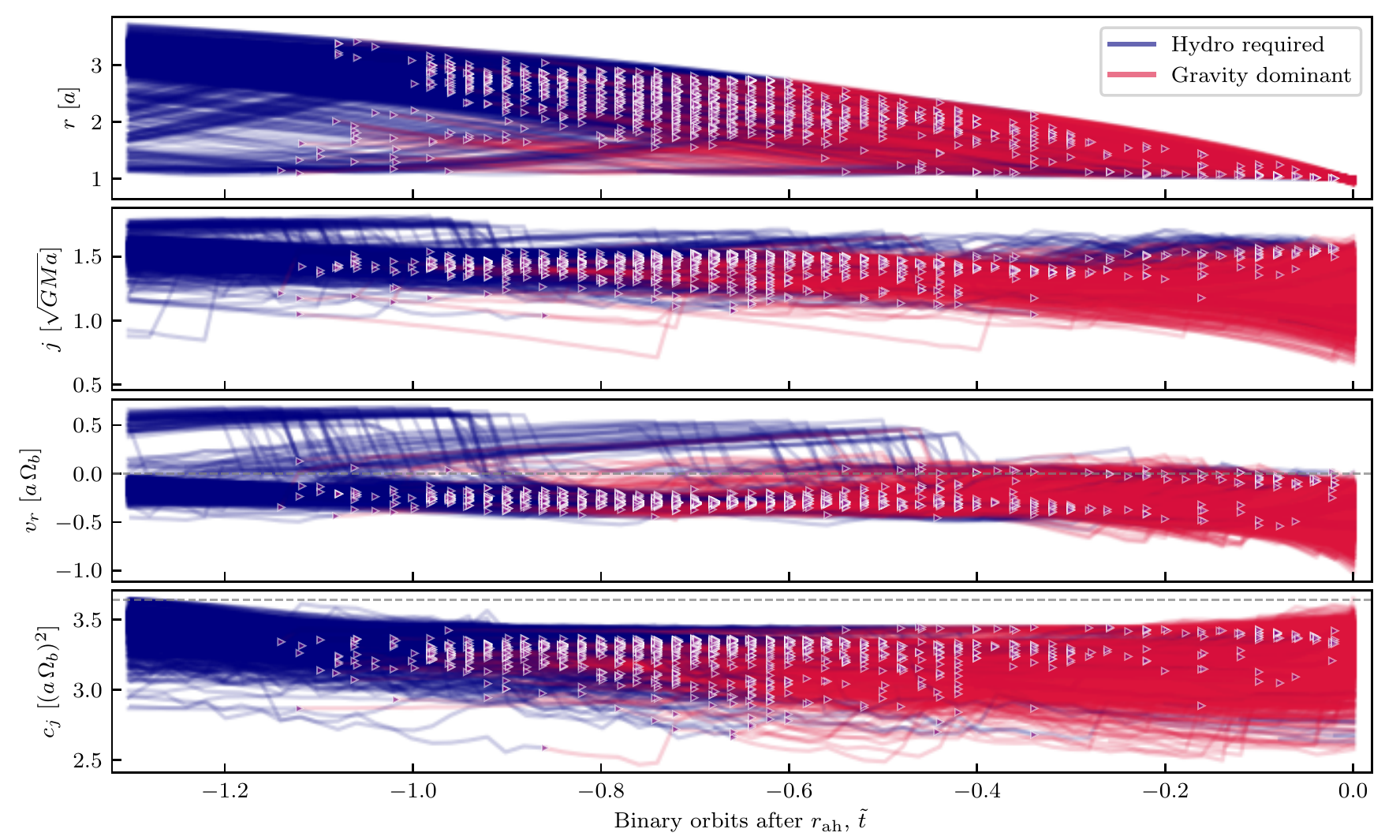}
\caption{Timeseries of radius ($r$), specific angular momentum ($j$), radial velocity ($v_r$), and Jacobi constant ($c_j$) for all tracers accreted between 650-660 orbits.  At each output time, the tracers are integrated in the cr3bp for 1.25 orbits from their phase-space coordinates: the tracer timeseries is plotted in blue when the given phase-space coordinates do not result in ballistic accretion, and switch to red when gravity becomes a full descriptor of tracer accretion.  The purple arrows with white outlines denote the time at which each fluid element's becomes gravitationally destined to accrete.}
\label{fig:histories}
\end{figure*}

Before turning towards what causes fluid elements to accrete, one question we sought to answer was at what point is a fluid element considered to be accreted, and as a corollary, what is the balance of flow across the binary cavity.  In particular, does fluid have to make it all the way to a minidisk in order to eventually accrete? And conversely, how much of the flung, or rejected, stream material is coming from fluid that has made it to (or even previously been incorporated into) a minidisk?

To examine this, we considered all tracers in a 10 orbit window ($650 - 660$ orbits) that started beyond some nominal radius $\bar r$ then crossed inside it ($r_0 > \bar r \rightarrow r < \bar r$) within the 10 orbits. We then looked at the distribution of tracer radii one binary orbit ($1 \, P_b$) after having crossed $\bar r$ ($r < \bar r \rightarrow r_1$) for 30 different $\bar r$ values in $0.7\, a < \bar r < 1.5 \, a$. Four of these distributions are illustrated in Figure \ref{fig:rbar} for $\bar r = [0.8\,a, 1.0\,a, 1.2\,a, 1.4\,a]$. 
We observe that as $\bar r$ decreases, the amount of material being flung back out to radii $r_1 \gtrsim 1.5\,a$ similarly decreases, and by $\bar r \lesssim 1.0\,a$, 99\% of all tracers that cross $\bar r$ are accreted and considered incorporated into minidisks ($r_1 < 1.0\,a$).  
We quantify this further in Figure \ref{fig:noreturn}  by integrating the distributions from the topmost panel in order to determine the empirical probabilities of accretion ($r_1 < 1.0\,a)$ and expulsion ($r_1 > 1.5\,a$), respectively, given that a tracer--or fluid element--has penetrated the radius $r < \bar r$.  
We observe, consistent with our intuition from Figure \ref{fig:rbar}, that accretion is a functional inevitability after having penetrated a radius $\bar r \lesssim 1.0\,a$, and the probability of expulsion only becomes non-negligible for $\bar r \gtrsim 1.1\,a$.  

The bottom panel of Figure \ref{fig:noreturn} shows the maximum radius achieved after a tracer has crossed $\bar r$ as determined by the 99th-percentile radius in the distributions from Figure \ref{fig:rbar}. 
From this we once again see that for small enough $\bar r$, all tracers that crossed were accreted, but as $\bar r$ is raised, the population of tracers that do escape back to the cavity wall following their close encounter re-emerges.
Because of the instability of orbits in the cavity region, the transition from a single peak distribution of purely accreted gas to a double peaked distribution that also contains expelled/rejected fluid elements is distinct and characterized by a sharp drop-off in ${\rm Max}(r_1)$. 
Accordingly, we identify this sharp drop-off (in our fiducial run--shown as solid lines) at $\bar r = r_{\rm ah} \approx 1.05\,a$ as the accretion horizon whereby functionally all material that crosses the horizon is destined to be accreted onto a minidisk.
We include the same results for $t_{\rm sink}$ in the slow-sink limit (dashed lines) since altering the amount of material in the minidisks could affect this result.  While the numerical value of the accretion horizon shifts by a few percent to $r \approx 1.01\,a$, the qualitative behavior and approximate location of the horizon remain.
As such, for the remaining analysis we will consider fluid elements as ``accreted'' once they have crossed the accretion horizon $r_{\rm ah} = 1.05\,a$ and will use this interchangeably with the process of a fluid element joining a minidisk.
The location of this accretion horizon is shown in the left panel of Figure \ref{fig:reliefs}.

It is worth mentioning that there exist a small set of outlier tracer particles ($<0.5\%$) that happen into phase-space coordinates allowing them to slingshot around the binary and through the central most regions of the domain without impacting and being consumed by a minidisk, thus escaping back to the cavity wall. 
However, we do not observe any tracers that are incorporated into a minidisk and are later dislodged and returned to the cavity wall.  Once fluid elements are subsumed by a minidisk, they are destined to stay there and eventually be accreted by the central component.

Recent hydrodynamical studies of thin circumbinary disks have shown that including the binary in the simulation domain and resolving the central regions of the accretion flow are important for determining certain quantities such as the net angular momentum transfer rate and the binary migration rate \citep[e.g.][]{Yike17, MML19}. 
However, many earlier studies excised the binary and central cavity ($r \lesssim a$) from the simulation domain \citep{MacFadyen2008, Shi+2012, Dorazio2013, Farris15, MML17}, yet their observations on the disk morphology and binary accretion rate (sans the variability) remain mostly accurate and consistent with the present understanding.
The presence of such an accretion horizon offers an explanation as to why these studies were reasonably accurate despite not resolving the central-most regions of the flow; namely, that material that crossed the excision horizon at $r \sim a$ was functionally disconnected from the outer-disk and would no longer affect fluid in the simulation domain.

\subsection{Lagrangian accretion histories} \label{sec:accretion-histories}

In order to examine how tracers that end up in a minidisk are able to accrete, we looked at histories of the tracer specific angular momentum $j$, radial velocity $v_r$ and Jacobi constant $c_j$. 
Figure \ref{fig:histories} shows these histories for all tracers that crossed the binary cavity and joined a minidisk over a 10 orbit window (650-660 orbits).
The time coordinate ($\tilde t$) is rescaled as the number of binary orbits after each tracer crossed the accretion horizon $r_{\rm ah} = 1.05\,a$.  Before addressing the color-coding, we notice that we can see two primary populations of fluid elements that were accreted. 
The first and most prominent group (making up the thickest blue band at the beginning of each timeseries) are the fluid elements that are accreted directly from the cavity wall.
These gas parcels come from radii characteristic of the cavity apoapse ($2.9\,a \lesssim r \lesssim 3.5\,a$) with specific angular momentum $j \sim 1.5 \sqrt{GMa}$, and slightly negative (inward) radial velocities characteristic of eliptical orbits going from apoapse to periapse.
The second population of accreted parcels are those that participated in a stream within the prior orbit and can be seen as the band of blue trajectories that increase their radii from $r \lesssim 2.5\,a$, begin with larger specific angular momenta $j \gtrsim 1.55\,\sqrt{GMa}$, and have comparatively large positive radial velocities.  
In all four timeseries we can see these fluid elements impact and become incorporated into the cavity wall as their specific angular momenta are  lowered to $j \sim 1.5\, \sqrt{GMa}$ and their velocities are redirected by the ram pressure in the cavity wall onto orbits approaching periapse.
Examples of this post-expulsion redirection can be seen in the trajectories in the second row of Figure \ref{fig:trajectories}.
We also see a smaller population of fluid elements that have recently participated in a stream but hover near a Lagrange point at $r \sim 1\,a$ that either eventually fall onto a binary component minidisk or get swept in by another stream; but nonetheless do eventually accrete within $\sim 1$ orbit.

The color coding of Figure \ref{fig:histories} denotes the fate of each particle in the cr3bp.
From their phase-space coordinates at each output time, we integrate each tracer forward for $1.25\,P_b$ in the purely gravitational problem. 
For each tracer at each output time, the fluid element is classified by whether or not its purely gravitational orbit takes it within the accretion horizon $r\sim 1.05\, a$ -- at which point it is regarded as accreted -- or not.  
When a tracer's accretion becomes a ballistic process, it is plotted in red; and when the purely gravitational trajectory is not sufficient for accretion, the history is shown in blue.  
The point at which a fluid element transitions to a trajectory that is well approximated by gravity alone (again in the boolean sense of accretion vs. no accretion) is demarcated by a purple arrow with a white outline.  
The observations of note here are that (a) most fluid elements that cross the binary cavity and accrete become gravitationally destined to do so somewhere between $-1.0\,P_b \lesssim \tilde t \lesssim -0.4\, P_b$ where $\tilde t$ is the time after accretion, and (b) there do not appear to be any strong clumpings of transition-points (purple triangles) where the majority of fluid elements are deflected onto accreting orbits;
or where a fluid element appears to lose significant angular momentum so as to fall into a strongly eccentric orbit.  
The tracers appear to predominantly be on eccentric orbits around the cavity and to smoothly, and seemingly at random, transition onto ballistic orbits destined to accrete. 

\begin{figure}[t!]
\includegraphics{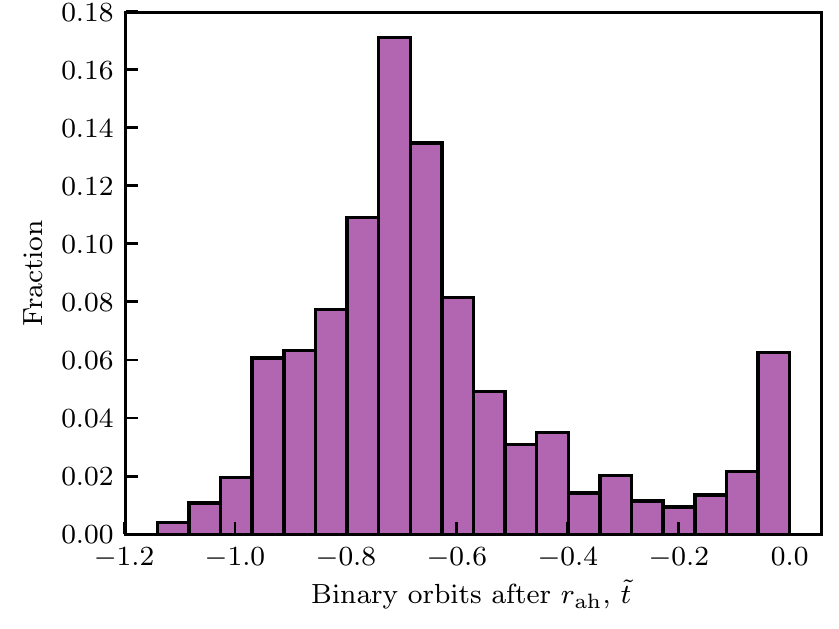}
\caption{
Histogram of the transition times (where fluid element accretion becomes a purely gravitational process) as denoted by the purple triangls in Figure \ref{fig:histories}.}
\label{fig:destinydist}
\end{figure}

\begin{figure*}[t!]
\includegraphics{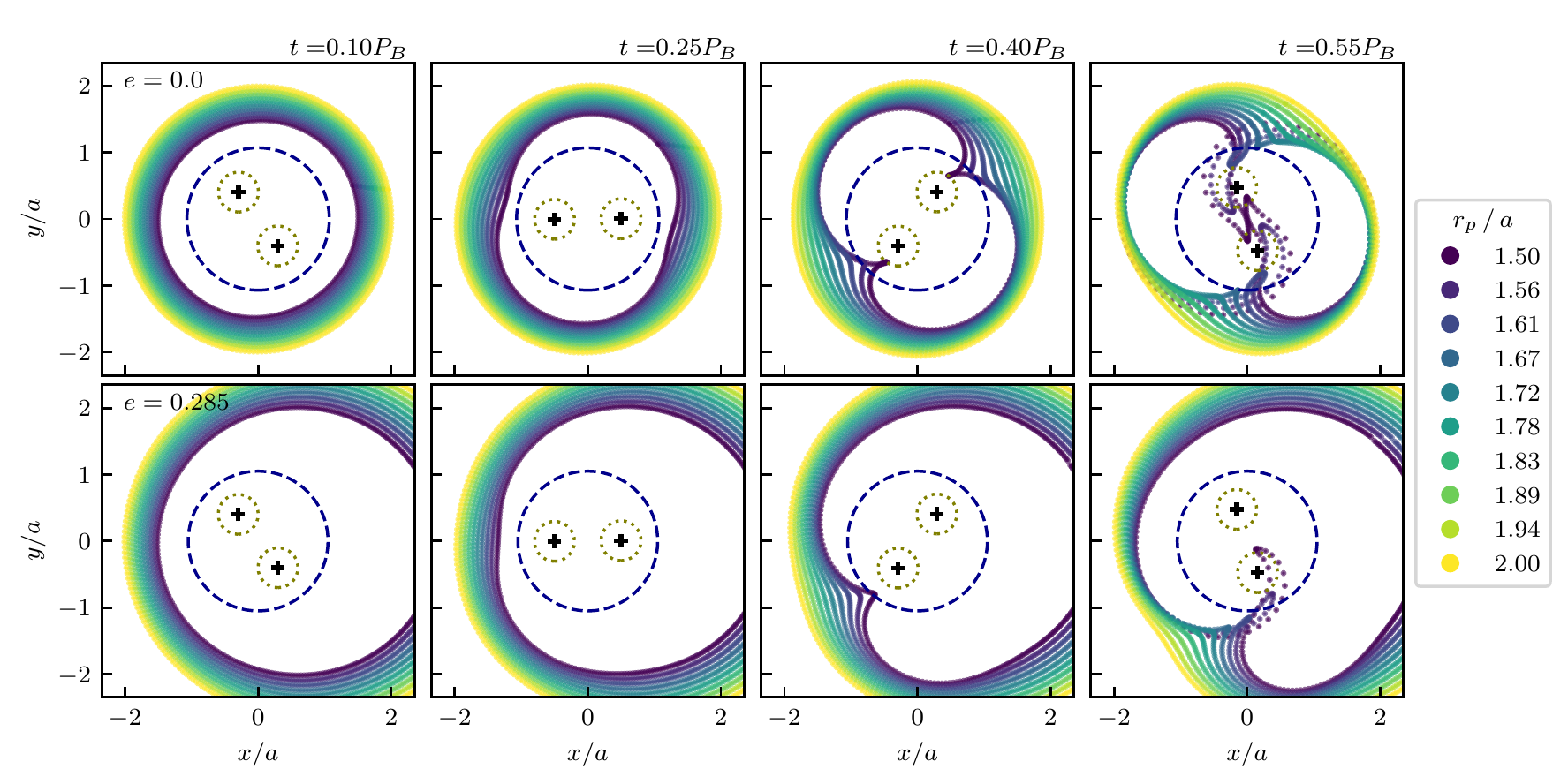}
\caption{
Tidal deformation of initially Keplerian orbits (prograde and counterclockwise).
The first row is for initially circular orbits, and the second row is for nested orbits with initial eccentricity of $e = \bar{e}_{\rm wall} \sim 0.285$. The binary is on a circular orbit in both cases, with both the binary and the fluid rotating counter-clockwise.
The radii listed in the legend denote the radius in the circular case and the radius of periapse in the eccentric case (such that the semi-major axis of said orbits is $a_p = r_p / (1 - e)$).}
\label{fig:cr3bp}
\end{figure*}

The distribution of times at which the accreted fluid elements become effectively ballistic is shown in Figure\,\ref{fig:destinydist}.  These times appear to approximately follow a gaussian distribution about $-0.7\,P_b$. Around $ 8\%$ of all accreted parcels only become gravitationally destined to do-so within the final moments ($\tilde t \lesssim -0.2\, P_b$) before crossing the horizon and being deposited onto a minidisk.  
These fluid elements that only enter the ballistic phase during the physical moments of stream formation are those that would not usually accrete in the cr3bp, but in the full hydrodynamic problem have their orbits slightly redirected by the transverse pressure-gradient as the stream resists orbit crossing (this population of ``pass-through'' tracers can be seen in the final panels of Figure \ref{fig:cr3bp}).

\subsection{Stream Formation}\label{sec:stream-formation}

To better understand the relative importance of gravitational forces and hydrodynamic forces in the transfer of fluid across the cavity, we performed a simple experiment of integrating nested rings of particles placed initially on Kepler orbits around a central binary \citep{Dorazio2013, Dorazio2016}.  
The results of this test are shown in Figure \ref{fig:cr3bp}.  
The top row shows the time evolution of 10 initially circular orbits through $0.55 \,P_b$.
The dashed blue circle shows the accretion horizon and the smaller tan dotted circles show
the approximate truncation radius, $0.3\,a$, of each minidisk \citep{al94, Eggleton83}.
Of primary interest, we note that the tidal perturbation of the initially circular orbits naturally results in the formation of streams that both cross the accretion horizon and also penetrate the Hill sphere of a binary component.  
The tendency of initially circular orbits to be deformed into streams due to gravity alone was similarly observed by \cite{Dorazio2013} \citep[see also][]{Pichardo2005, Pichardo2008}.
We find that circular orbits of radius $r \lesssim 1.61\, a$ result in Hill sphere penetrating tidal perturbations and $r \lesssim 1.72\,a$ give orbits that are perturbed beyond the accretion horizon.

The second row of Figure \ref{fig:cr3bp} shows the same experiment for nested initially-Keplerian orbits of eccentricity $e=0.285$ with their longitude of periapse initially perpendicular to the binary semi-major axis. 
We choose $e=0.285$ by taking the average of the eccentricities
\begin{align}
    e = \left| \frac{v^2\,\boldsymbol r - (\boldsymbol v \cdot \boldsymbol r)\,\boldsymbol v}{GM} - \boldsymbol{\hat r} \right|
\end{align}
of all tracer particles in the cavity wall.\footnote{We find that the cavity wall is well described as those particles with specific angular momenta $1.4\,\sqrt{GMa} < j < 1.8\,\sqrt{GMa}$.}
In the case of eccentric orbits the color represents the initial orbital periapse, $r_p$, such that the orbital semi-major axis can be determined as $a_p = r_p / (1 - e)$.  
The addition of eccentricity to our initially Keplerian rings breaks the symmetry of the tidal perturbations, but nonetheless, we observe the natural formation of stream structures that both cross the accretion horizon and deliver material within a component Hill sphere.  
The tidal deformation is slightly less prominent in the eccentric case because the orbital velocity at periapse exceeds that of the circular orbits, but we determine that orbits with $e=0.285$ and $r_p \lesssim 1.56\,a$ cross a binary Hill sphere and $r_p \lesssim 1.61\,a$ cross the accretion horizon.

We conclude that the formation of accretion streams and the deposition of fluid onto binary minidisks is a natural consequence of the tidal deformation of Keplerian orbits given that such orbits pass sufficiently close to the binary.  
Therefore, for a full fluid disk, so long as material can reliably be moved down to orbits with periapse passage $r_p \lesssim 1.6\,a$, the capture of fluid appears an entirely gravitational process.
As such, we define a tidal capture radius
$r = r_{\rm cap} = 1.6\,a$ such that initially Keplerian orbits with periapse radius $r_p < r_{\rm cap}$ will be tidally deformed into streams that penetrate the accretion horizon. $r_{\rm cap}$ is shown in Figure \ref{fig:reliefs}, and we see that there is in fact material within this radius at cavity periapse in the full hydrodynamic problem.

\vfill\null
\vfill\null
\vfill\null
\subsection{Minidisk Capture} \label{sec:minidisk-capture}
\begin{figure}[t!]
\includegraphics[width=\columnwidth]{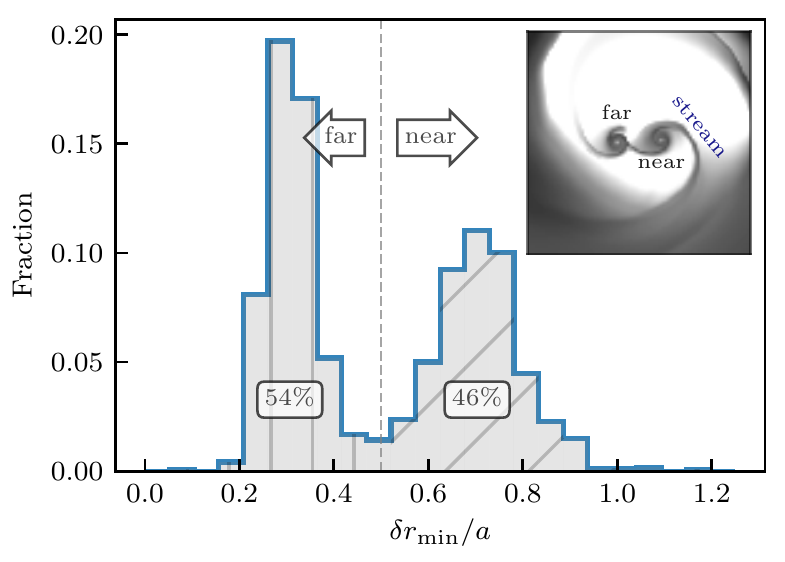}
\caption{
For each tracer accreted between 650-600 orbits, the closest approach to its non-accretor component. The left peak represents those fluid elements that swing around the near component and join the minidisk of the far component; and the right peak consists of those fluid elements that transfer directly to the near minidisk. 
The size of the minidisks is $\sim 0.3\,a$.
Approximately $\sim 46\%$ (diagonal hatching) of all accreted material transfers directly to the near minidisk, and $\sim 54\%$ (vertical hatching) joins the minidisk of the far component.
The inset shows an accretion stream with its respective ``near'' and ``far'' components labelled for illustration.}
\label{fig:capture}
\end{figure}
\begin{figure}[t!]
\includegraphics[width=\columnwidth]{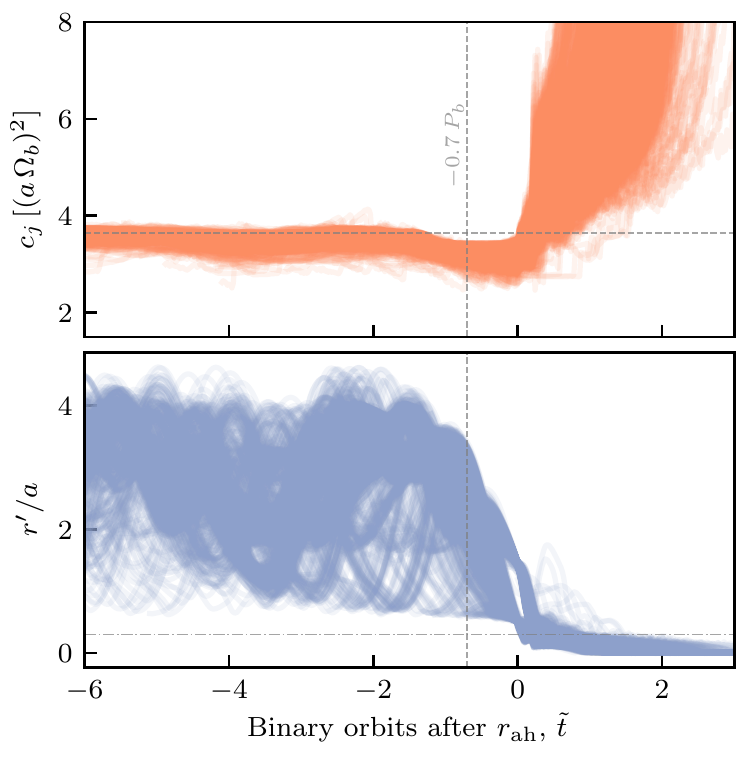}
\caption{
(Top) Timeseries of Jacobi constants for all accreted tracers from 650-660 orbits. The dashed horizontal line denotes $c_j^{\rm crit}$. 
(Bottom) Distance from each tracer's ``accretor'' component.  The dotted dashed line shows the truncation radius of the minidisk.
The vertical line shows the approximate peak time where elements become ballistic accretors from Figure \ref{fig:destinydist}.
}
\label{fig:cjacc}
\end{figure}

The approximately ballistic gas streams fall from the CBD wall toward one of the minidisks. Some gas parcels are subsumed directly into this ``near'' minidisk, while others skirt its outer edge and get transferred to the ``far'' minidisk. The BH onto which the parcel ultimately accretes (the ``accretor'') can be either the near or far one. 
By measuring the closest approach $\delta r_{\rm min}$ a gas parcel makes to its non-accretor BH, we can characterize the capture process in terms of the fraction of gas which falls directly from the CBD wall onto its accreting BH. 
Parcels which do not experience a close approach to the non-accreting component ($\delta r_{\rm min} \gtrsim r_{\rm Hill}$) are said to have accreted directly. 
Figure \ref{fig:capture} shows the distribution of $\delta r_{\rm min}$. It is bimodal, indicating that a comparable amount of gas accretes directly to the near component, as accretes indirectly to the far one. Accretion to the far component is marginally favored (54\%).

Moreover, we see the effect of the accretion shock in the top panel of Figure \ref{fig:cjacc}.  
As fluid elements are deformed into a stream and approach the accretion horizon, their Jacobi constant has dipped below $c^{\rm crit}_j$ (a necessary--but not sufficient--condition for ballistic mass transfer); but when they cross the horizon and impact a minidisk, $c_j$ increases rapidly, crosses $c_j^{\rm crit}$, and they become bound to their respective minidisk. The distance from the component each fluid element becomes bound to ($r^\prime$) and the minidisk truncation radius are shown in the bottom panel for reference.

\subsection{CBD Loss Cone} \label{sec:loss-cone}
\begin{figure}[t!]
\includegraphics{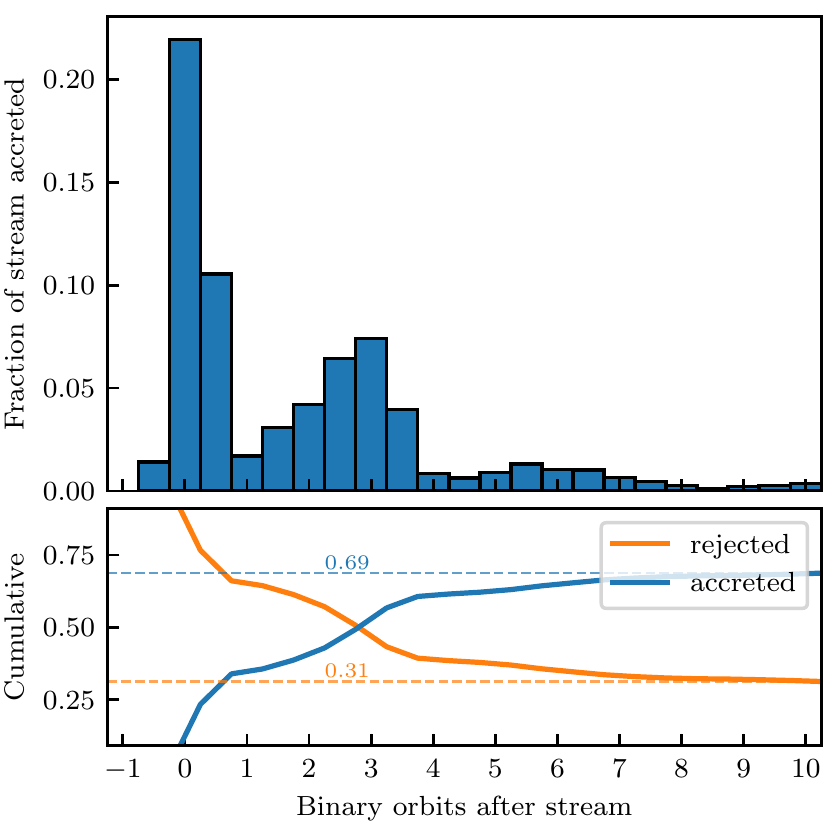}
\caption{Histogram of tracer times-to-accrete after having been pulled into a stream taken over a 10 orbit window (650-670 orbits).  While most fluid elements ($\sim 30\%$) are accreted in the orbit that spawned the stream, there is a second accretion spike $\sim 3$ orbits later.  After $\sim 6$ orbits, the remaining $\sim 30\%$ of fluid elements appear to be well-mixed in the cavity wall, with little-to-no memory of their "close-passage" in a stream.
}
\label{fig:streams}
\end{figure}

The presence of the accretion horizon and the fact that stream formation and mass transfer onto minidisks is a predominantly gravitational process poses similarities to the theory of loss-cone orbits. 
The traditional loss-cone, for objects in orbit around a massive binary, is defined as those orbits with small enough angular momentum at given energy $E$, $L(E) \leq L_{\ell c}(E)$ such that their impact parameter is within $b < \gamma\,a$;
where gamma is some factor $\gamma \sim \mathcal{O}(1)$ defining the ``slingshot'' radius of the central binary (or in the single-object case, the disruption radius); \citep{fpp2003}.  
For orbits with $r \gg \gamma a$, or equivalently when $|E| \ll GM / \gamma a$, the critical angular momentum defining the loss cone is $L_{\ell c}(E) = 2(\gamma a)^2 \left[ E - \phi_g(\gamma a) \right] \approx 2 G M \gamma a$.  
In the scenario of a binary accreting from a thin disk of fluid (or nested Keplerian rings), we have determined such a capture radius inside of which the accretion of fluid elements is gravitational, $r_{\rm cap} = \gamma_{\rm cap} \, a \approx 1.6\, a$.

For stellar mass objects in orbit around a massive binary, those that pass close to the binary are imbued with angular momentum, removed from the loss cone, and slingshot to larger radii.  
However, for the CBD loss cone, the presence of a viscous disk of fluid prevents such gravitational slingshots from sending the higher angular momentum material to large radii.  
When a stream is formed, $\sim 70\%$ of the stream material (see Figure \ref{fig:streams}) is imparted with some additional angular momentum and is flung back to the CBD (visible in the trajectories of Figure \ref{fig:histories} starting with $j \gtrsim 1.55 \sqrt{GMa}$).  
In the absence of the CBD, this material would be removed from the binary's radius of influence, but it instead impacts the cavity wall and is immediately redirected onto orbits once again eligible for tidal capture.

This process is evident in Figure \ref{fig:streams} which shows how many orbits after participating in a stream it takes a fluid element to accrete.
The bottom panel shows the cumulative fraction of stream-tracers accreted after each orbit.  
The time of the stream is taken as each time the binary semi-major axis is perpendicular to the cavity longitude of periapse (which is assumed constant since the cavity precession time is $\mathcal{O}(10^2\, P_b)$), and data is taken from 650-670 orbits (equivalent to 40 streams).  
We see that $\sim 30\%$ of the tracers comprising these streams are directly accreted within the orbit that spawned the stream.  
There is a second accretion spike 2-3 orbits after the initial stream as some of the fluid elements that were originally expelled to the cavity wall have been redirected onto orbits where they once again are swept into a stream and deposited onto a minidisk.  
This second accretion spike could be evidence for the shock deflection suggested by \cite{Shi2015}; but it could also be a result of the fact that the flung material is fanned across the far side of the cavity (see Figure\,\ref{fig:reliefs}) such that some of it has an orbital commensurability with the moments of stream formation 2-3 orbits later. 
Approximately 4 orbits after the formation of a stream, $\sim 60\%$ of its fluid elements have been accreted, and the remaining material has been well mixed back into the cavity wall and has completely forgotten any history of having participated in a stream.

We note that an important element of this study is that the gravitational slingshot is necessarily interrupted by the cavity wall because it is confined to the plane of the disk.  However, in 3D it could be that some of the material is redirected out of the disk-plane and could possibly escape the system in out-of-the plane,
binary-torque driven ``winds''. We intend to quantify this in future work.

Returning to this notion of the CBD loss cone, while the rate at which the loss cone is refilled in the traditional problem is set by the two-body relaxation time, in the case of a circumbinary disk this timescale is set by the viscous time in the inner disk. 

\begin{figure}[t!]
\includegraphics{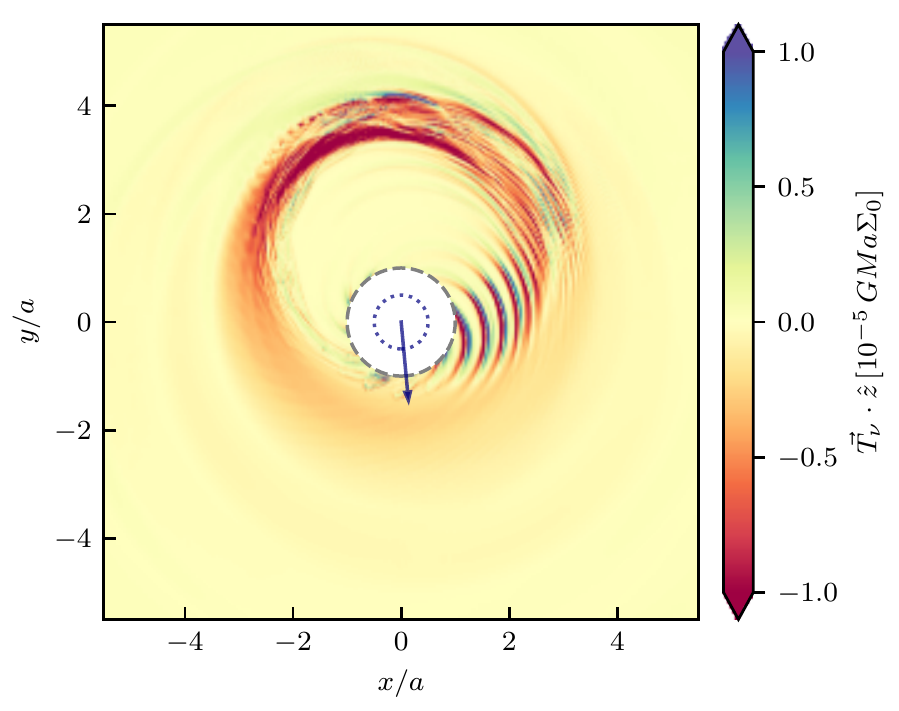}
\caption{Map of viscous torques averaged over 10 orbits in the lab frame (treating the cavity orientation as fixed).  The blue arrow shows the average cavity argument of periapse and the iner dotted circle shows the binary orbit.  We see that viscous torques are maximal and negative near cavity apoapse.}
\label{fig:viscmaps}
\end{figure}

In Figure \ref{fig:viscmaps} we show 10 orbit averaged maps of the viscous torque for the inner disk in the observer frame.  
As in Figure \ref{fig:trajectories}, we have excised the inner most $r < a$ for visual clarity.
The dark blue arrow denotes the average longitude of periapse.
We see that the torques are maximal at cavity apoapse, and that the predominant effect is to remove angular momentum from fluid elements in the cavity wall. 
In this way, as fluid elements orbit the binary, viscous drag through the majority of the orbit extracts angular momentum, moving the inner most orbits onto increasingly eccentric orbits with decreasing $r_p$.  
Thus, at some point in their last orbit around the inner edge of the cavity, gas parcels lose enough angular momentum that they become gravitationally destined for tidal capture.
We posit that it is for this reason that no clear indicator appears in Figure \ref{fig:histories} marking the imminence of accretion, and why the transition from ``requiring hydrodynamics for accretion'' to ``ballistically destined to accrete'' is seemingly random.

\subsection{Implications for long-term evolution} \label{sec:implications}

The notion of mass transfer from the CBD on to minidisks as the result of delivering material close enough ($r < \gamma_{\rm cap} \, a$) to the binary for tidal capture provides a natural mechanism by which the CBD can regulate its accretion rate.  
Namely, one can imagine two characteristic accretion rates: (1) the viscous feeding rate in the outer disk $\dot M_{\rm cbd}$, and (2) the rate at which mass is tidally stripped off the inner edge of the cavity wall and actually fed onto minidisks $\dot M_{\rm cav}$. There is no ab-initio reason for these two accretion rates to be the same. However, in the case of an infinite disk in a true steady state where the amount of material accreted through the binary ($\dot M$) matches the outer feeding rate $\dot M = \dot M_{\rm cbd}$, it must also be that $\dot M_{\rm cbd} = \dot M_{\rm cav}$.  
In the absence of a true steady-state, since $\dot M_{\rm cav}$ is set by the inner edge of the cavity wall, if $\dot M_{\rm cav} > \dot M_{\rm cbd}$ the binary will scour away the inner edge of the cavity wall, material will not be delivered fast enough to replenish the wall and remove angular momentum from the inner most orbits, the inner edge will recede such that $r_p > \gamma_{\rm cav}\, a$, and $\dot M_{\rm cap}$ will decrease or turn off completely. Conversely, if $\dot M_{\rm cbd} > \dot M_{\rm cav}$ mass will pile up into the lump at the cavity wall, the lump will grow increasingly dense and steep until the viscous torque begins to infringe upon the centrifugal barrier, the cavity inner edge will shift nearer to the binary increasing the prominence of tidal deformations and the mass of accretion streams, and $\dot M_{\rm cav}$ will increase.
Such effects have been observed in \cite{Rafikov2016}.
For a sufficiently relaxed disk, then, these two effects will balance yielding a steady (in the time-averaged sense) cavity structure that mediates the equivalence $\dot M_{\rm cbd} = \dot M_{\rm cav} = \dot M$.

\begin{figure}[t!]
\includegraphics{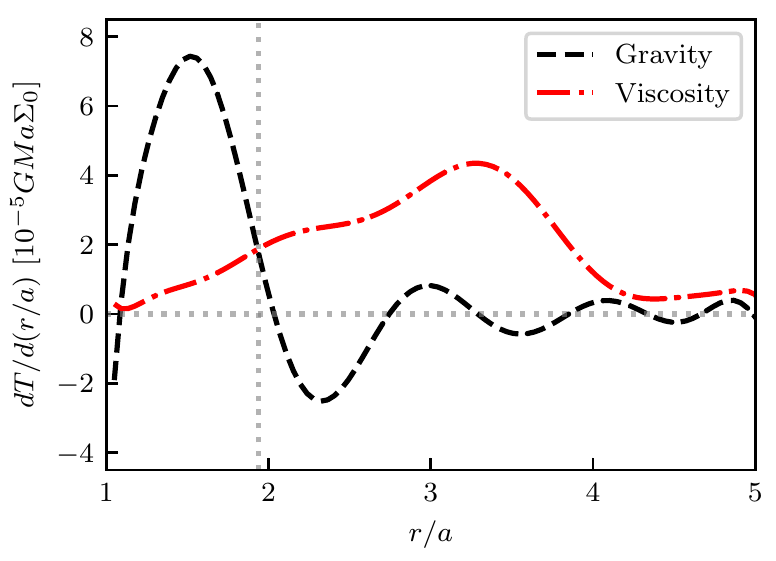}
\caption{Time-averaged radial profiles of gravitational and negative-viscous torque densities. The radius at which the two are equal, at $r = 1.95\,a$, is shown with a vertical dotted line and corresponds with the approximate location of the cavity wall.}
\label{fig:torque-balance}
\end{figure}

Moreover, to understand the location of the inner edge of the binary cavity, in Figure \ref{fig:torque-balance} we show the time-averaged radial profiles of the gravitational and negative viscous torque densities in the disk
\begin{align}
    \frac{d T_{\rm g}}{dr} &= \,\,\,\, 2\pi r \left\langle \Sigma\, \frac{d\phi_g}{d\phi} \right\rangle_{\phi,\, t} \\
    -\frac{d T_{\rm \nu}}{dr} &= -2\pi \left\langle \frac{d}{dr}\bigg( \nu \, r^3 \big\langle \Sigma\, \partial_r\Omega \big\rangle_\phi \bigg) \right\rangle_t \ .
\end{align}
The analytic form of the binary potential derivative is used and the rest of the quantities are measured from the simulation checkpoints over 10 orbits \citep[e.g.][]{MacFadyen2008, Cuadra2009, Roedig2012, Shi+2012, Dorazio2013, Rafikov2016}. 
The approximate, axisymmetric location of the cavity edge is given by the balance
\begin{align}
    \frac{d T_{\rm g}}{dr} = -\frac{d T_{\rm \nu}}{dr} \, ,
\end{align}
and is shown by the dotted vertical line in Figure \ref{fig:torque-balance}.  
In our simulations, this torque balance occurs at $r = 1.95\, a$ which visually agrees with the approximate average cavity locations seen in Figures\,\ref{fig:reliefs} and \ref{fig:viscmaps}.  
We can think of this axisymmetric approximation to the cavity location as the average cavity distance seen by the binary in the co-orbitting frame.
Moreover, this balance implies that one can move the location of the cavity wall slightly by varying the disk viscosity (but only slightly because the gravitational torque density profile near the equilibrium is largely inelastic).  
Increasing the viscosity would cause the viscous torque profile to shift upwards and move the cavity edge closer to the binary, and decreasing $\nu$ would shift it downwards, moving the cavity further away.  
This effect has been observed in studies that vary the disk viscosity \citep[e.g.][]{MML17, Ragusa2020}.  
In the limit of zero viscosity $\nu = 0$, we see that a cavity would still form at $r \sim 2\,a$ due to the root in the gravitational torque density, consistent with notions of non-intersecting stable orbits at $r \sim 2\,a$ \citep{Pichardo2005, Pichardo2008}.  
In this way, we posit that binary accretion is the result of viscosity moving material 
from the outer disk to the cavity edge, where--if near enough--orbits are tidally deformed and captured as ballistic accretion streams.
We note that this average cavity radius as seen by the binary is not sufficient to induce tidal capture and mass transfer across the cavity, and the instantaneous cavity eccentricity ($e\approx 0.3$ as measured in Section \ref{sec:stream-formation}) is necessary for delivering orbiting fluid elements beyond the tidal capture radius.

%

%
% =============================================================================
\section{Summary and Conclusions} \label{sec:summary}

We have studied the histories of how fluid elements accrete in a two-dimensional isothermal disk around an equal-mass, circular binary via tracer particles embedded in Eulerian hydrodynamics.  Our primary findings can be summarized as follows:
\begin{enumerate}
    \item[(i)] There exists an accretion horizon---a radius beyond which functionally no material is returned to the cavity wall---at $r \approx 1.05\,a$ for binaries embedded in thin, isothermal circumbinary disks (Figure \ref{fig:noreturn}).
    \item[(ii)] Nearly all accreted fluid elements become gravitationally destined to do so around $\sim 0.7$ orbits before crossing the accretion horizon.  Moreover, this ballistic transition does not appear to correlate with abrupt changes in angular momentum, radial velocity, or Jacobi constant. This suggest that strong shocks are not the primary mechanism to transport angular momentum and allow gas to accrete.
    \item[(iii)] We demonstrate that stream formation and mass transfer onto minidisks is driven by a gravitational process.  Accordingly, we determine a tidal capture radius, $r_{\rm cap} \approx 1.6\,a$.
\end{enumerate}
In accordance with the observations (i)-(iii), we develop a description of the mechanism by which fluid elements are accreted from thin circumbinary disks around circular equal-mass binaries and draw parallels to the theory of loss-cone orbits. Specifically, the accretion of fluid from the outer CBD follows a three-stage evolution: 
\begin{enumerate}
    \item[(1)] Fluid elements are viscously transported to the inner-CBD and cavity wall. The stresses grow as the element moves through the lump and closer to the cavity wall.
    \item[(2)] Gas parcels persist in the inner-regions of the CBD and cavity wall ebbing and flowing through the binary's quadrupolar potential until they fall onto periapse passages with $r_p < r_{\rm cap}$ \emph{and} the proper azimuthal phase to be tidally captured in accretion streams.
    \item[(3)] Fluid elements are subsumed into a minidisk via an accretion shock.
\end{enumerate}
At this point the element is bound to the minidisk and fated for accretion onto its binary component.

There are a number of simplifying assumptions we made that could influence these results. 
We have employed a locally isothermal equation of state, but a more thorough treatment of the disk thermodynamics as well as the radiation could meaningfully alter the flow. 
Similarly, we have restricted ourselves to two-dimensions and ignored both magnetic fields and general relativistic effects. 
The other primary limitation of this work is that we have only considered equal-mass binaries fixed on circular orbits.  
Moreover, this picture is tailored to large mass-ratio binaries and does not apply to gap-carving binaries in the planetary regime $q \lesssim 0.025$.
We might naively expect our picture to apply for more eccentric orbits and for binaries of differing mass-ratios (and $q \gtrsim 0.1$), but this is left to future study.  

%
% =============================================================================
\acknowledgments

We thank the anonymous referee for a constructive report and Mulin Ding for administering the Ria computing cluster at NYU.  Resources supporting this work were provided by the NASA High-End Computing (HEC) Program through the NASA Advanced Supercomputing (NAS) Division at Ames Research Center. 
We acknowledge support by NSF grants AST-2006176 (to ZH) and 1715661 (to ZH and AM).

\bibliographystyle{mnras}
\bibliography{refs}

%
% =============================================================================
\section*{Appendix A: Tracer Tests}

In this appendix we present a selection of idealized tests in order to probe the reliability of our tracer particle implementation (Equation \ref{eq:tracers}). First, we measure the ability of the tracers to track the flow and angular momentum in a steady-state disk solution.  
Then, we examine two tests on the ability of the tracers to accurately follow the flow of mass in our simulations; the significance being that if the tracers can accurately follow the mass flow in the disk, then all other instantaneous hydrodynamic quantities can be queried at all points in time in order to recreate the hydrodynamic history of a given mass parcel in a Lagrangian picture of flow.

\subsection*{Steady-State Disk}

\begin{figure}[h!]
\centering
\includegraphics{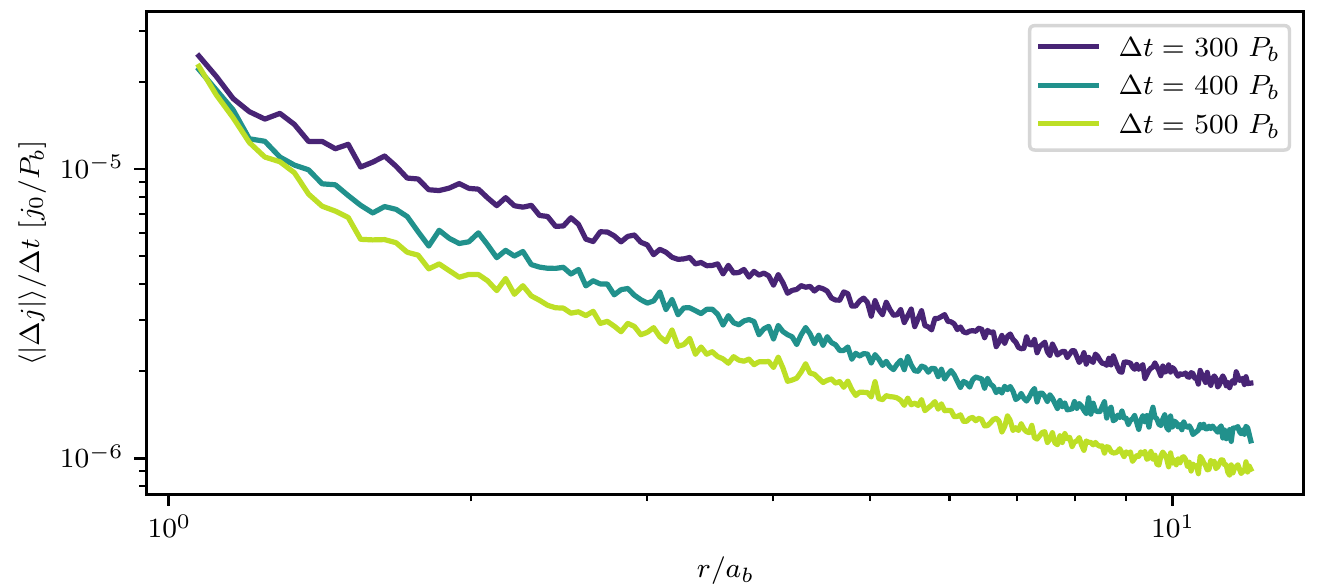}
\caption{Average percent change in tracer specific angular momentum, $\langle \Delta j \rangle / j_0$, per equivalent-binary-orbit ($P_b$) in an inviscid disk around a single central mass.  It would require $>100$ binary orbits to change a tracer's $j$ by $\sim 1\%$ at $r = 1\,a_b$.}
\label{apx:disk-delta-j}
\end{figure}

As a first test, in order to quantify the degree of diffusion in the tracer implementation we consider the steady-state form of the initial condition presented in Section \ref{sec:setup}: the disk of Equations \ref{eq:sigma0} and \ref{eq:v0} around a single point-mass of mass $M$ with viscosity $\nu = 0$. In this way, the radius and angular momentum of all fluid elements in the disk will only be changed by numerical dissipation introduced at the grid scale.  
We run this steady-state disk for a numerical time equivalent to 500 binary orbits (with binary separation $a_b$) and measure the average percent change in tracer specific angular momentum, $\langle \Delta j \rangle / j_0$ (with $j_0$ the initial specific angular momentum), per equivalent-binary-orbit ($P_b$).
This is shown in Figure \ref{apx:disk-delta-j} after 300, 400, and 500 equivalent-binary-orbits.
While we find that the tracers are slightly more dissipative in the inner-most regions of the steady-state solution, it requires more than 100 equivalent-binary orbits for the most dissipative average tracer to change its specific angular momentum by 1\%. 
This $1\%$ error-time in the viscous disk ($\nu = 0.001$) would be longer than a local viscous time.

\subsection*{Spreading Ring}
\begin{figure}[h!]
\includegraphics{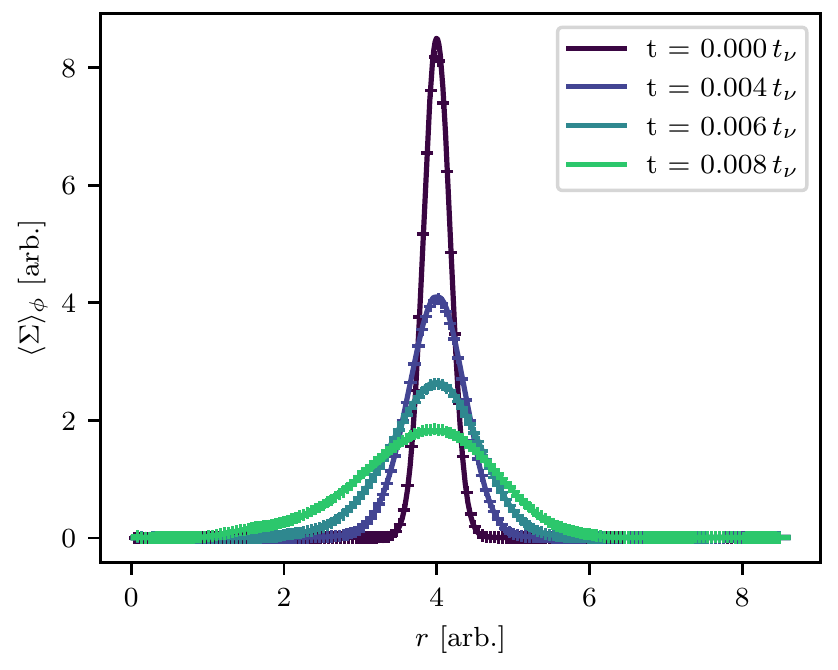}
\includegraphics{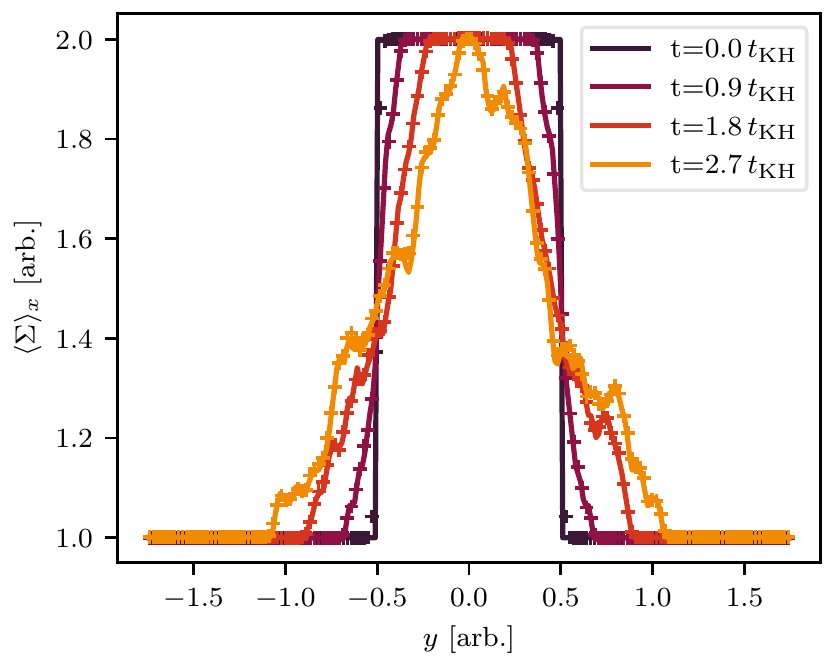}
\caption{Average density profiles in the gas (lines) and calculated from the tracer distributions (crosses) for (left) a spreading Keplerian ring, and (right) a Kelvin-Helmholtz Instability (see Figure \ref{apx:khi-tracers}).}
\label{apx:ring-test}
\end{figure}

In order to test the tracers' ability to track the redistribution of mass in our simulations, we ran a 2D spreading ring test for a gaussian ring of width $\sigma=0.25$ initially centered at $r_0=4$ (in arbitrary units of length) such that
    \begin{align*}
        \Sigma = \Sigma_0 \, \textrm{exp} \left[ - \left(\frac{r - r_0}{\sigma}\right)^2 \right] \ .
    \end{align*}
The total ring mass set by $\Sigma_0$ is also arbitrary, the initial velocity profile is Keplerian, and the kinematic viscosity is set to $\nu=0.01$.
The evolution of the disk's radial surface density profile measured in the Eulerian fluid (lines) and in the tracer particle distribution (crosses) is shown in the left panel of Figure \ref{apx:ring-test}. The simulation was performed with uniform spatial resolution $\Delta = 0.0104$ in a domain extended from $[-8, 8]$ in both the x- and y-direction with arbitrary length units.  
We initialize $\sim 5 \times 10^5$ tracers evenly distributed on the grid where each tracer is assigned a ``weight'' defining the amount of mass associated with its assigned Lagrangian fluid element.  
The sum of all weights is equivalent to the total mass in the domain. 
Times are reported in units of the viscous time at the initial ring location $r_0$, and we see that the tracers follow the spreading of the ring extremely accurately.
We note that the spreading of the initially gaussian ring is not entirely viscous as the simulation is run with an isothermal equation of state (Equation \ref{eq:eos}), but the purpose of the test is not to examine the accuracy of the code's viscosity prescription; but rather to inspect the ability of our tracer particles to accurately follow the mass evolution of the system. In this test, the tracers accurately follow the diffusion of the gas.

\subsection*{Kelvin-Helmholtz Instability}
\begin{figure*}[h!]
\includegraphics{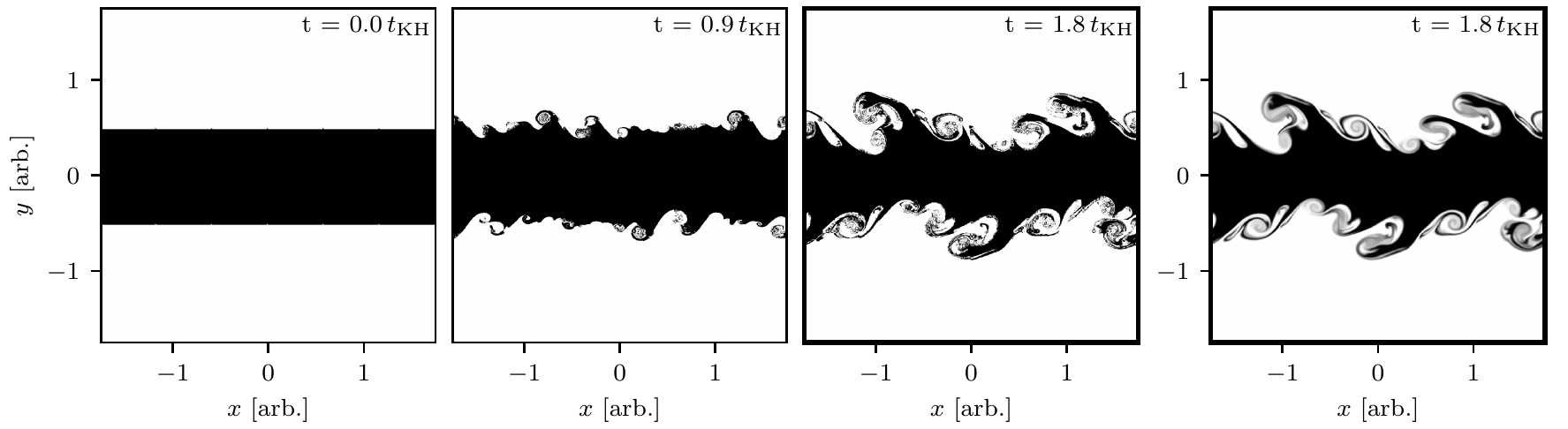}
\caption{(First 3 panels) All tracers from the middle high-density region as the KHI grows. (Last panel) Gas density of the fluid at the same time as the final panel of tracer distributions.}
\label{apx:khi-tracers}
\end{figure*}

We additionally ran a classical Kelvin-Helmholtz Instability (KHI)
test in order to quantify the tracer particle accuracy at the onset of turbulent flow with significant mixing layers and rapid accelerations (as opposed to the nearly laminar coherent motions of a steady or slowly spreading disk).  
In this problem we evolve the energy equation with an ideal gas law equation of state and $\gamma = 5/3$. An HLLC approximate Riemann solver is also employed \citep{Toro} in order to accurately preserve the initially shearing contact discontinuities.  
The simulation is performed in a 2D periodic box of extent $\Delta x = 3.5$ in each direction (again with arbitrary length units). 
We initialize the fluid with a high-density strip $\Sigma_1 = 2.0$ and right-flowing velocity $v^{(x)}_1 = 0.1$ for $-0.5 < y < 0.5$. The rest of the domain is initialized with a lower density $\Sigma_2 = 1.0$ and a left flowing velocity $v^{(x)}_2 = -0.1$. 
The pressure is initially uniform.  
The instability is seeded by a superposition of damped, sinusoidal vertical velocity perturbations $v_y = v_0\,\cos{(k\,x)}\, (e_+ + e_-)$ with $v_0 = 0.002$, $e_\pm = \exp{[-0.5 * (y \pm 0.5)^2]}$, and $k = \pi$.
The system is seeded uniformly with $\sim 5 \times 10^5$ tracer particles with their weights accounting for the mass of their Lagrangian fluid elements. 

Figure \ref{apx:khi-tracers} shows the evolution of the KHI in the tracer particles (first 3 panels) initially placed in the high-density strip (black dots). 
We see the emergence of increasingly large ``rolls'' as the flow evolves.  
The timescale for the growth of the instability is given by $t_{\rm KH} = 2\pi  \sqrt{\Sigma_2 / \Sigma_1}\, (k\,\Delta v)^{-1}$ where $k$ is the wavenumber of the velocity perturbation and $\Delta v = v_1^{(x)} - v_2^{(x)}$.
For comparison, the final panel shows the gas density at the same time as the bolded panel of tracer distributions. We see that the spiral arms and mixing layers are recreated quite precisely in the tracer distributions.  
More quantitatively, the right panel Figure of
\ref{apx:ring-test}
shows the $x$-averaged linear density profiles in both the gas (lines) and constructed from the tracer distributions (crosses). We see that the tracer particles are able to follow the vortices and gas mixing throughout the onset of the KHI to a high degree of accuracy.

\end{document}